\documentclass{article}

\usepackage{arxiv}

\usepackage[utf8]{inputenc} 
\usepackage[T1]{fontenc}    
\usepackage{hyperref}       
\usepackage{url}            
\usepackage{booktabs}       
\usepackage{amsfonts}       
\usepackage{nicefrac}       
\usepackage{microtype}      
\usepackage{lipsum}
\usepackage[fontsize=12pt]{scrextend}

\usepackage{graphicx}
\usepackage{amssymb}
\usepackage{array}
\usepackage{wasysym}
\usepackage[citestyle=numeric-comp,bibstyle=ieee,language=auto,babel=other,%
backend=biber,sorting=none,doi=false,maxnames=3]{biblatex}
\renewbibmacro*{bbx:savehash}{}

\newcommand{\be}{\begin{equation}}
\newcommand{\ee}{\end{equation}}
\newcommand{\beqa}{\begin{eqnarray}}
\newcommand{\eeqa}{\end{eqnarray}}
\newcommand{\dd}{\,{\rm d}}

\title{The Orbital Lense--Thirring Precession\\ in a Strong Field}

\author{Vladimir N. Strokov\thanks{\texttt{vnstrokov@gmail.com} (corresponding author)}\,\,\,$^{\textstyle a,b}$ \\ 
   \And
 Shant Khlghatyan\,$^{\textstyle b}$ \\
 \And \\
$^{\textstyle a}$\,Lebedev Physical Institute, Astro Space Centre \\ 
84/32 ul. Profsoyuznaya, Moscow, Russia 117997 \\
  \And \\
$^{\textstyle b}$\,Moscow Institute of Physics and Technology \\
9 Institutskiy per., Dolgoprudny, Moscow Region, Russia 141701 \\
}

\begin{document}
\maketitle

\begin{abstract}
We study the exact evolution of the orbital angular momentum of a massive particle in the gravitational field of a Kerr black hole. We show analytically that, for a wide class of orbits, the angular momentum's hodograph is always close to a circle. This applies to both bounded and unbounded orbits that do not end up in the black hole. Deviations from the circular shape do not exceed~$\approx10\%$ and~$\approx7\%$ for bounded and unbounded orbits, respectively. We also find that nutation provides an accurate approximation for those deviations, which fits the exact curve within~$\sim 0.01\%$ for the orbits of maximal deviation. Remarkably, the more the deviation, the better the nutation approximates it. Thus, we demonstrate that the orbital Lense--Thirring precession, originally obtained in the weak-field limit, is also a valid description in the general case of (almost) arbitrary exact orbits. As a by-product, we also derive the parameters of unstable spherical timelike orbits as a function of their radii and arbitrary rotation parameter~$a$ and Carter's constant~$Q$. We verify our results numerically for all the kinds of orbits studied.
\end{abstract}


\section{Introduction}\label{intro}

Solutions to the two-body problem in Newtonian celestial mechanics are well known and exhaustively described by two properties: the orbits remain in one plane and fall in one of the three classes: ellipse, parabola, or hyperbola. As general relativity (GR) comes into play, those properties are violated (except for special cases). In the gravitational field of a Schwarzschild black hole, although the orbits are still confined to a plane, their shape becomes increasingly complicated as dimensionless parameter $GM/(r_{\rm m}c^2)$ grows ($G$ is the gravitational constant, $c$ is the speed of light; $M$, the mass of the attracting center, and $r_{\rm m}$ is a minimal distance to that center). When the parameter is small, for bounded orbits the approximation of a precessing ellipse works well (e.g.~\cite{landau} or see~\cite{Shapiro_2011} for derivation in terms of the Laplace--Runge--Lenz vector). However, in the fully relativistic regime this approximation is not valid anymore.

In the field of a rotating Kerr black hole, nor remain the orbits in a single plane (one exception is equatorial orbits). We say that a particle's orbit is confined to one plane if the particle's total angular momentum~$\mathbf{L}\equiv (L_x,L_y,L_z)$ is constant, projections of the angular momentum being
\beqa
	L_{x}&=&p_{\theta}\sin\phi+p_{\phi}\cos\phi\cot\theta\,, \label{eq:Lx} \\
	L_{y}&=&p_{\phi}\sin\phi\cot\theta-p_{\theta}\cos\phi\,, \label{eq:Ly} \\
	L_{z}&=&p_{\phi}\,, \label{eq:Lz}
\eeqa
where $\phi$ and $\theta$ are the angles of either the conventional curvature coordinates of the Schwarzschild metric~\cite{MTW,frolov_novikov}, or Boyer--Lindquist coordinates of the Kerr metric~\cite{Visser}, and $p_{\phi}$\,, $p_{\theta}$ are the respective covariant components of the particle's momentum. This is a convenient representation of an apparent shape of the orbit plotted against ``Cartesian'' axes~$(x,y,z)$ which are related to the angles~$\theta, \phi$ and to Schwarzschild or Boyer--Lindquist radius~$r$ through the conventional formulae of a transformation to spherical coordinates\,\footnote{From the viewpoint of intrinsic geometry, the plane is, of course, not flat and has a nonvanishing curvature. Also note that the ``Cartesian'' coordinates are different from the Kerr--Schild coordinates.}. Although the components~$L_x$, $L_y$, and~$L_z$ do not come from any Killing vector in the Kerr space--time, they do quantify the measurable effect of the change of a particle's orbital plane with respect to a distant observer. As we are to show, the evolution of this plane in terms of these components is quite simple (though the derivation of the result is rather lengthy).

\begin{figure}[!htbp]
\begin{center}
	\includegraphics[width=0.6\textwidth]{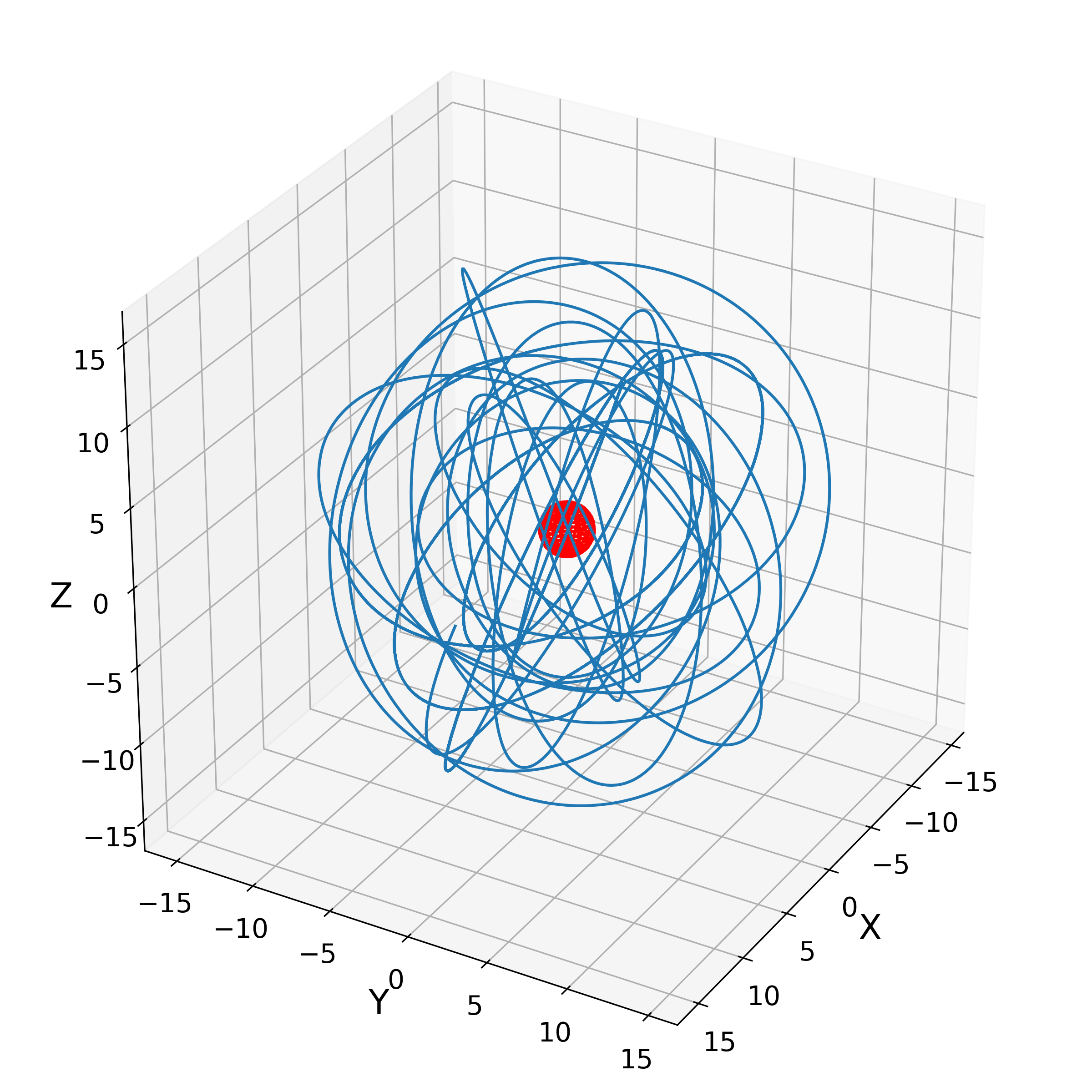}
\end{center}
	\caption{An off-plane orbit of a massive particle around a black hole with rotation parameter $a=0.85$. The particle starts from $r_0 = 12$, $\theta_0 = \pi/2$, and $\phi_0 = 0$ and moves in the direction of increasing~$\theta$ and decreasing $r$ and $\phi$. The constants of motion are $E=0.963$, $L_z = -1.37548$, and $Q = 14.0$ (see Sec.~\ref{review} for notation).}
	\label{intricate_orbit}       
\end{figure}

Nonetheless, as long as the Kerr's field is weak enough, that out-of-the-plane motion can be attributed to the Lense--Thirring precession of the angular momentum~\cite{LT_original,L-T}, that is, of the orbital plane\,\footnote{The precession of the axis of a gyroscope~\cite{Schiff_1960} which is also referred to as the Lense--Thirring effect is not pertinent to this Paper (see also a discussion of certain analogies between the two effects in Conclusion).}. When the field is fully relativistic, the orbits can be quite intricate (see Figure~\ref{intricate_orbit} for an example) and, in particular, essentially spherical~\cite{Wilkins_1972,Goldstein_1974,Teo_2003}. Also, a number of rigorous results on geodesics in Lorentz manifolds were obtained by means of geometric analysis, e.g.~\cite{Giannoni_1991,Masiello_1992,Hasse_2006}. On the other hand,  a visual reduction of these seemingly unarranged orbits to a superposition of simpler motions would contribute to intuitive understanding of the celestial mechanics in a Kerr black hole's gravitational field.


In this Paper, we propose such a reduction and show that any orbit of a massive particle that does not end up in the black hole results from the superposition of a motion in a plane and the precession and nutation of that plane. Although, strictly speaking, this reduction is not exact, the superposition approximates the orbit within~$\sim 0.01\%$. In order to show it, we study the exact evolution of orbital angular momentum and then confirm our results numerically. We find that for most orbits the precession alone approximates the exact motion reasonably well. The discrepancy can be further accounted for by introducing nutation, the ``second-order precession'' around a precessing axis. And as it turns out, the larger the discrepancy, the better the nutation accounts for it. The superposition of precession and nutation is reminiscent of Ptolemaic epicycles, but remarkable is the fact that there is no need to consider higher-order precessions. Therefore, unlike the orbits themselves, their orbital momenta demonstrate a relatively simple behavior. 


\begin{table}[!htbp]
\centering\caption{Major notation.\label{notation}}
\begin{tabular}{>{\centering\arraybackslash}p{0.3\columnwidth}p{0.7\columnwidth}}
\hline\noalign{\smallskip}
Notation & Explanation  \\
\noalign{\smallskip}\hline\noalign{\smallskip} 
$(t,r,\theta,\phi)$ & the Boyer--Lindquist coordinates of the Kerr metric  \\
$M$ & the mass of a Kerr black hole \\
$a$ & the angular momentum of a Kerr black hole (in the units of~$M$) \\
$
\begin{array}{rcl}
\Delta &\equiv&  r^2 -2r + a^2 \\
\rho^2 &\equiv& r^2 + a^2\cos^{2}{\theta} \\
\end{array}
$ & frequently used combinations \\
$E$ & the conserved energy of a particle (in the units of the particle's rest mass) \\
$L_x, L_y, L_z$ & component of the orbital angular momentum of a particle \\
$Q$ & Carter's constant \\[0.3cm]
$\dot{\left(\ldots\right)}\equiv \frac{\dd\left(\ldots\right)}{\dd\tau}$ & derivative with respect to proper time~$\tau$ of a particle \\[0.3cm]
$\gamma$: $\dd\gamma = E\dd\tau/\rho^2$ & auxiliary parameter along geodesics \\[0.3cm]
$\mathcal{L}\equiv \sqrt{L_x^2 + L_y^2}$ & radius of an angular momentum's hodograph in $(L_x,L_y)$-plane \\[0.3cm]
$\Delta Q \equiv a^2(E^2 - 1)\cos^{2}{\theta}$ & variation of~$\mathcal{L}^{2}$ \\[0.3cm]
$r_{\rm isco}^{\pm}$, $r_{\rm mb}^{\pm}$, $r_{\rm ph}^{\pm}$ & radii of innermost stable circular orbit, marginally stable orbit, and photon orbit, respectively (plus and minus stand for co- and counterrotating orbits, correspondingly) \\[0.3cm]
$
\begin{array}{rcl}
y &\equiv& 1/r \\
y_{\pm} &\equiv& 1/r_{\rm ph}^{\pm} \\
q &\equiv& 1/Q \\
\epsilon &\equiv& E/\sqrt{Q}\\
\lambda_z &\equiv& L_z/\sqrt{Q}\\
x&\equiv& \lambda_z - a\epsilon
\end{array}
$ & auxiliary notation \\
$y_0$ & inverse radius of a spherical orbit with $L_z - aE = x\sqrt{Q} =0$ \\
$y_{*}$ & inverse radius of a spherical orbit with $L_z = 0$ in the limit $Q\to +\infty$ \\ 
\hline\noalign{\smallskip}
\end{tabular}
\end{table}


The Paper is organized as follows. In Sec.~\ref{review} we review the geodesic equations of the Kerr space--time and provide a formula for the radius of a hodograph of the orbital angular momentum (recall that a hodograph is the locus of the end of a vector). Bounded and unbounded orbits are considered in Sec.~\ref{bounded} and Sec.~\ref{unbounded}, respectively. In Sec.~\ref{nutation} we introduce nutation to account for deviations of the hodographs from a circle and estimate its magnitude in Sec.~\ref{observe_nut}. In Sec.~\ref{conclusion} we discuss our results.

In what follows, the signature~$(-\,+\,+\,+)$ and units $G=c=1$ are used. All lengths and times are in units of~$M$, the mass parameter of a black hole. Major notations are summarized in Table~\ref{notation} while minor notations are given along the text.

\section{Geodesic equations}\label{review}

Consider the Kerr space--time in Boyer--Lindquist coordinates, e.g.~\cite{Visser}:
\beqa
\dd s^2&=&-\left(1-\frac{2r}{\rho^2}\right)\dd t^2 +\left(r^2+a^2+\frac{2ra^2}{\rho^2}\sin^2\theta\right)\sin^2\theta \dd\phi^2 -\nonumber \\
&-&\frac{4ra}{\rho^2}\sin^2\theta \dd\phi \dd t +\frac{\rho^2}{\Delta}\dd r^2+\rho^2\dd\theta^2 \,.
\eeqa

As is known~\cite{chandrasekhar2}, a massive particle's free fall in this space--time is characterized by a few constants of motion involving covariant components of 4-momentum \mbox{$p_\mu \equiv (p_t,p_r,p_\theta,p_\phi)$}. Those are energy at infinity \mbox{$E=-p_{t}$} conventionally measured in units of the particle's rest mass, a component of the orbital angular momentum~$L_z=p_{\phi}$ that is parallel to the rotation axis of the black hole, and the so-called Carter's constant~\cite{Carter_1968} defined as
\be
\label{eq:Q}
Q=p_{\theta}^2+\cos^2\theta\left[a^2(1-E^2)+\frac{L_z^2}{\sin^2\theta}\right]\,.
\ee
In addition, conserved is the norm of the 4-momentum vector, $p_{\mu}p^{\mu}=-1$.

Given a set of parameters~$(E,L_{z},Q, a)$, geodesic equations can be partially integrated to yield~\cite{chandrasekhar2}:
\beqa
&\rho^4\dot{r}^2&=R(r)\equiv (E^2-1)r^4+2r^3+[a^2(E^2-1)-L_z^2-Q]r^2 +  \nonumber \\ 
& &\qquad\qquad +2[(aE-L_z)^2+Q]r-a^2Q\,, \label{eq:geodes1} \\
&\rho^4\dot{\theta}^2&=\Theta(\theta)\equiv Q-\cos^2\theta\left[a^2(1-E^2)+\frac{L_z^2}{\sin^2\theta}\right]\,,\label{eq:geodes2} \\
&\rho^2\dot{\phi}&=\frac{1}{\Delta}\left[2arE+(\rho^2-2r)\frac{L_z}{\sin^2\theta}\right]\,, \label{eq:geodes3}\\
&\rho^2\dot{t} &= a(L_z-aE\sin^2\theta)+\frac{r^2+a^2}{\Delta}[E(r^2+a^2)-aL_z]\,, \label{eq:geodes4}
\eeqa
where the dot stands for a derivative with respective to proper time~$\tau$.

In this form, there is a partial reduction of the particle's motion into independent evolution along $r$ and $\theta$. In the $\theta$-direction, the particle is moving in angular potential
\be
U_{\theta}(\theta,L_z,E) = \cos^2\theta\left[a^2(1-E^2)+\frac{L_z^2}{\sin^2\theta}\right]\,,
\ee
with the Carter's constant~$Q$ playing the role of ``angular energy''. Also, since function~$R(r)$ can be factorized~\cite{Wilkins_1972} as follows\,\footnote{Another factorization proposed in~\cite{Krivenko_1976} can be convenient if the sum $Q + (L_z - aE)^2$ is used instead of the Carter's constant.}:
\be
R(r)=rS(E-U_{-})(E-U_{+})\,, 
\ee
with
\beqa
\label{radial_pot}
U_{\pm}(r,L_z,Q)\equiv\frac{2aL_z \pm \sqrt{\Delta\left[r^2L_z^2+(r + Q/r)S\right]}}{S}\,, \nonumber \\
S\equiv r^3 + a^2 r + 2a^2\,,
\eeqa
the particle's motion in the $r$-direction occurs in the radial potential~$U_{+}(r,L_z,Q)$.


Since $L_z$ is an integral of motion, the evolution of the angular momentum is completely given by the latter's hodograph in the $(L_x,L_y)$-plane. Consider the squared radius of this hodograph, which, from~(\ref{eq:Lx})--(\ref{eq:Lz}) and~(\ref{eq:Q}), reads:
\be
\label{eq:L2}
\mathcal{L}^2\equiv L_x^2 + L_y^2 = Q +\Delta Q\,, \qquad \Delta Q = a^2(E^2-1)\cos^{2}{\theta}\,.
\ee
Note that the first term in the right-hand side is constant. Therefore, $\mathcal{L}$ varies only due to the second term~$\Delta Q$, and the ratio of the two terms determines whether the hodograph will deviate from a circle significantly.

We proceed by separately considering bounded and unbounded orbits and assume that neither of them enter the black hole.

\section{Bounded orbits}\label{bounded}

In this case, $E<1$ and $Q\geq 0$ (see~\cite{Wilkins_1972} and eq.~(\ref{eq:Q})). Hence, $|\Delta Q|< 1$ and, if $Q>>1$, a hodograph's radius $\mathcal{L}$ is approximately constant, i.e. a circle is a good approximation for the hodograph.

If $Q<<1$, $Q$ and $|\Delta Q|$ may be comparable. Indeed, as long as $E<1$, $\theta=\pi/2$ is a stable equilibrium point~\cite{Krivenko_1976} of angular potential~$U_\theta$. Therefore, for a given~$Q$, the maximal value of~$\cos^{2}{\theta}$ is on the order of~$Q$, see~(\ref{eq:geodes2}). Namely,
\be
\left.\cos^{2}{\theta}\right|_{\rm max} =  \frac{Q}{L_z^2 + a^2(1-E^2)} + \mathcal{O}\left(Q^2\right)\,, \quad Q\to 0\,, 
\ee
such that
\be
\label{bounded_equator}
\frac{|\Delta Q|}{Q} = \frac{a^2(1-E^2)}{L_z^2 + a^2(1-E^2)} + \mathcal{O}\left(Q\right) \lesssim 1\,,
\ee
where $L_z$ and $E$ are the parameters of equatorial orbits~$Q=0$. 

Seemingly, in the last case $|\Delta Q|$ could be equal to~$Q$ when~$L_z =0$ (even exactly, according to~(\ref{eq:geodes2})). However, this is not realized as long as the orbit is bounded. Moreover, in order to estimate the r.h.s. of~(\ref{bounded_equator}) from above, it is sufficient to consider orbits of constant~$r$ (and equatorial ones in the case in question), because, for a given~$L_z$, to maximize the r.h.s., $E$ must be as low as possible, i.e. equal to the minimum of~$U_{+}(r)$.

Recall that the parameters of equatorial circular orbits are given by relations~\cite{Bardeen_1972,circorb0}:
\beqa
%
E(y,a)&=&\frac{1-2y\pm ay^{3/2}}{(1-3y\pm 2ay^{3/2})^{1/2}}\,,  \label{eq:Ecirc}\\
L_{z}(y,a)&=&\pm \frac{1+a^2 y^2\mp 2ay^{3/2}}{\left[y(1-3y\pm 2ay^{3/2})\right]^{1/2}}\,, \label{eq:Lcirc}
\eeqa
where $y\equiv 1/r$, $r$ being the radius of a circular orbit, and the upper and lower signs stand for co- and counterrotation, correspondingly.

These orbits are divided into the following subclasses:
\begin{itemize}
\item $y\in(0,1/r_{\rm isco}^{\pm})$, $r_{\rm isco}^{\pm}$ is the radius of the Innermost Stable Circular Orbit (hereafter, ``$\pm$'' in the index also stand for co- and counterrotation, respectively); orbits are stable,

\item $y\in(1/r_{\rm isco}^{\pm},1/r_{\rm mb}^{\pm})$, $r_{\rm mb}^{\pm}$ is the radius of the Marginally Bound orbit; orbit are unstable with $E<1$ (an inward perturbation sends the particle into the black hole and an outward perturbation accompanied by a decrease in energy sends the particle into a noncircular bounded orbit),

\item $y\in(1/r_{\rm mb}^{\pm},1/r_{\rm ph}^{\pm})$, $r_{\rm ph}^{\pm}$ is the radius of the circular PHoton orbit; orbits are unstable with $E>1$ (an inward perturbation sends the particle into the black hole and, under an outward perturbation, the particle escapes to infinity).
\end{itemize}

The characteristic radii are ordered, $r_{\rm isco}^{\pm}\geq r_{\rm mb}^{\pm}\geq r_{\rm ph}^{\pm}$, the equalities holding for an extremely rotation black hole, $a=1$, \textit{and} corotating orbits, and read:
\beqa
r_{\rm isco}^{\pm}&=&3+Z_2\mp\sqrt{(3-Z_1)(3+Z_1+2Z_2)}\,,\\
Z_1&=&1+(1-a^2)^{1/3}[(1+a)^{1/3}+(1-a)^{1/3}]\,, \nonumber \\
Z_2&=& \sqrt{3 a^{2}+Z_{1}^{2}}\,, \nonumber \\
r_{\rm mb}^{\pm}&=& 2\mp a + 2(1\mp a)^{1/2}\,, \\
r_{\rm ph}^{\pm}&=& 2\left[1+\cos{\left(\frac 23 \arccos{(\mp a)}\right)}\right]\,.
\eeqa

Note that both $E(y,a)$ and~$L_z^2(y,a)$ as functions of~$y$ have minima at~$y=1/r_{\rm isco}^{\pm}$ (see Appendix~\ref{EL_minimum}). Therefore, the leading term in the r.h.s. of~(\ref{bounded_equator}) has a maximum at that point, because
\be
\frac{a^2(1-E^2)}{L_z^2 + a^2(1-E^2)} \equiv 1 - \frac{1}{1 + a^2(1-E^2)/L_z^2}
\ee
and $(1-E^2)>0$ for bounded orbits as well as $L_z^2>0$. Also, the value of this maximum grows as~$a\to 1$ and is the highest for corotating orbits as~$r_{\rm isco}^{+}\to 1$. In that limit~~\cite{Bardeen_1972}, $E\to 1/\sqrt{3}$ and~$L_z\to 2/\sqrt{3}$, and from~(\ref{eq:L2}) and~(\ref{bounded_equator}) we obtain:
\be
\label{bounded_max}
\frac{|\Delta Q|_{\rm max^{+}}}{Q} = \frac 13 \quad \Rightarrow \quad \frac{\mathcal{L}_{\rm max} - \mathcal{L}_{\rm min}}{\mathcal{L}_{\rm max} + \mathcal{L}_{\rm min}} = \left(\sqrt{3}-\sqrt{2}\right)^2 \approx 0.10\,.
\ee

For counterrotating orbits, $E\to 5\sqrt{3}/9$, $L_z\to -22\sqrt{3}/9$, and
\be
\frac{|\Delta Q|_{\rm max^{-}}}{Q} = \frac{1}{243} \quad \Rightarrow \quad \frac{\mathcal{L}_{\rm max} - \mathcal{L}_{\rm min}}{\mathcal{L}_{\rm max} + \mathcal{L}_{\rm min}}  \approx 0.0010\,.
\ee

\begin{figure*}
\begin{center}
	\includegraphics[width=0.6\textwidth]{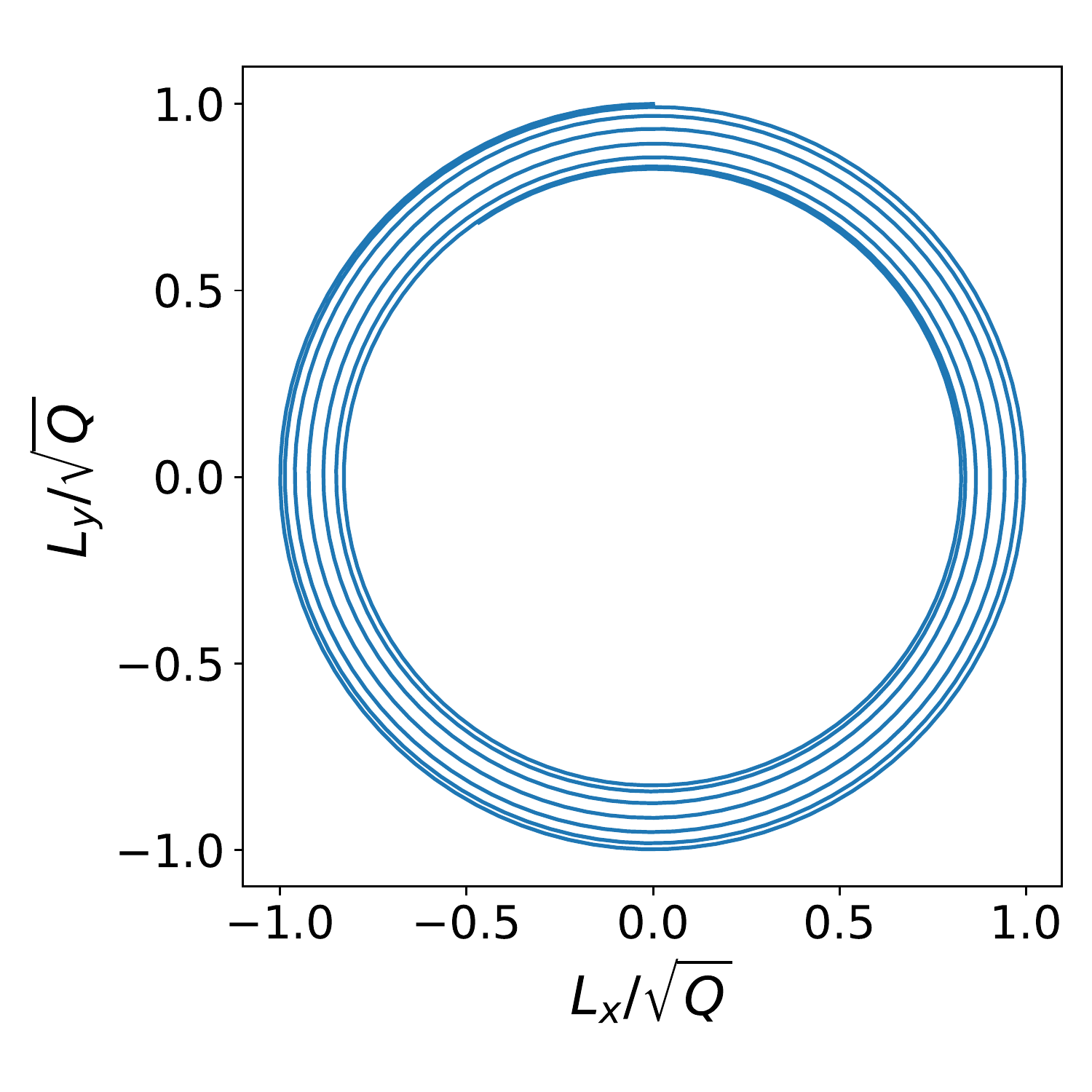}
\end{center}
	\caption{A maximally noncircular hodograph of the orbital angular momentum of a massive particle orbiting an extreme black hole, $a=1$. The particle starts from $r_0 = 1.03$, $\theta_0 = \pi/2$, and $\phi_0 = 0$. The constants of motion are $E=0.59$, $L_z = 1.18026$, and $Q = 0.00437$.}
	\label{superrelativistic_bounded}       
\end{figure*}

Thus, the hodographs of bounded orbits deviate from a circle by no more than \mbox{$\approx 10\%$}. Figure~\ref{superrelativistic_bounded} shows such a close-to-extreme hodograph.


\section{Unbounded orbits}\label{unbounded}

A particle on an unbounded orbit that escapes to infinity after having been reflected by radial potential~$U_{+}$ has $E\geq 1$. Therefore, there are two possibilities for~$Q$ and~$\Delta Q$ to be comparable. The first possibility is similar to the case~$Q<<1$ for bounded orbits. The second possibility arises, because one might expect~$\Delta Q>>Q$ if~$E>>1$. Let us consider these possibilities in more detail.

\underline{$Q<<1$}. In the range between~$1/r_{\rm mb}^{\pm}$ and~$1/r_{\rm ph}^{\pm}$, function~$a^2(E^2(y,a)-1)$ is less than function~$L_z^2(y,a)$, their ratio being maximal at~$y=1/r_{\rm ph}^{\pm}$ (notice that either function tends to infinity at that point). Therefore, $\theta=\pi/2$ continues to be a stable center of the angular potential~$U_{\theta}(\theta)$.

Also, that maximal value is the highest for an extremely rotating black hole. In the case of corotation~\cite{Bardeen_1972}, $(E/L_z)^2\to 1/4$ at~$y=1/r_{\rm ph}^{+}$ as~$a\to 1$. Hence, similarly to~(\ref{bounded_max}),
\beqa
\frac{\Delta Q_{\rm max^{+}}}{Q}& = & \frac{a^2(E^2-1)}{L_z^2 - a^2(E^2-1)}=\frac 13 \quad \Rightarrow \nonumber \\
\nonumber \\
&\Rightarrow& \qquad \frac{\mathcal{L}_{\rm max} - \mathcal{L}_{\rm min}}{\mathcal{L}_{\rm max} + \mathcal{L}_{\rm min}} = \left(2-\sqrt{3}\right)^2 \approx 0.07\,.
\eeqa
For counterrotating orbits, $(E/L_z)^2\to 1/49$ at~$y=1/r_{\rm ph}^{-}$ as~$a\to 1$. Therefore,
\beqa
\frac{\Delta Q_{\rm max^{-}}}{Q}& =& \frac{a^2(E^2-1)}{L_z^2 - a^2(E^2-1)}=\frac{1}{48} \quad \Rightarrow \nonumber \\
\nonumber \\
&\Rightarrow& \qquad \frac{\mathcal{L}_{\rm max} - \mathcal{L}_{\rm min}}{\mathcal{L}_{\rm max} + \mathcal{L}_{\rm min}} = \left(2-\sqrt{3}\right)^4 \approx 0.005\,.
\eeqa

\underline{$Q>>1$}. Consider also $E>>1$, because only in this case may we expect~$\Delta Q\sim Q$. Hence, $\Delta Q/Q \lesssim a^2 E^2/Q$\,.

On the other hand, the upper limit for~$E$ is the energy of an unstable spherical orbit, corresponding to a maximum of potential~$U_{+}(r)$, eq.~(\ref{radial_pot}), or a minimum of function~$R(r)$, eq.~(\ref{eq:geodes1}). To find the latter, we divide~$R(r)$ by~$Q$ and neglect terms~$\propto 1/Q$. Then, the conditions for a nonstable spherical orbit read:
\beqa
R_{Q}(y)&\equiv& \epsilon^{2} - 2 \epsilon x\cdot a y^{2}  -  x^{2}y^{2}(1  - 2 y) - y^{2}(1 - 2 y + a^{2} y^{2}) = 0\,, \label{RQ}\\
\frac 12 \frac{\dd R_{Q}}{\dd y} &=& - 2 \epsilon x \cdot a y  - x^{2}y(1- 3 y)  
- y (1 - 3 y + 2 a^{2} y^{2}) = 0\,, \label{RQdiff}\\
\frac 12 \frac{\dd^2 R_{Q}}{\dd y^2} &=& - 2 \epsilon x\cdot a  - x^{2}(1 - 6 y) - (1 - 6 y + 6 a^{2} y^{2}) > 0\,,
\eeqa
where
\be
\epsilon\equiv \frac{E}{\sqrt{Q}}\,, \qquad x \equiv \lambda_z - a\epsilon\,, \qquad \lambda_z(y,a) \equiv \frac{L_z}{\sqrt{Q}}\,.
\ee

Solving the second equation for~$\epsilon$ and substituting to the first equation, we find
\be
x = - \frac{2 a^{2} y^{2} - 3 y + 1}{\sqrt{4 a^{2} y^{3} - 9 y^{2} + 6 y - 1}}\,, \qquad 1/r_{\rm ph}^{-} < y < 1/r_{\rm ph}^{+}\,. \label{xq0}
\ee
Note that the numerator is proportional to a derivative of the expression inside the square root. Also, recall that $y_{\pm}\equiv 1/r^{\pm}_{\rm ph}$ are solutions to equation $1-3y\pm 2ay^{3/2}=0$, i.e. they are roots of the denominator.

Other variables are:
\beqa
\epsilon(y,a) &=&  \frac{a y^{2} \left(1 - y \right)}{\sqrt{4 a^{2} y^{3} - 9 y^{2} + 6 y - 1}}\,, \label{epsq0}\\
\lambda_z(y,a) &=& -\frac{1 - 3 y + a^{2} y^{2} + a^{2} y^{3} }{\sqrt{4 a^{2} y^{3} - 9 y^{2} + 6 y - 1}}\,, \label{lambdaq0} \\
\frac{\dd^2 R_{Q}}{\dd y^2} &=&  \frac{8 a^{2} y^{2} \left[ (1-y)^3 + y^3 (1-a^2)\right]}{\sqrt{4 a^{2} y^{3} - 9 y^{2} + 6 y - 1}} > 0\,.
\eeqa

Notice that, if $0<a<1$, $\epsilon$ and~$\lambda_z$ diverge as~$y\to y_{\pm}$ while their ratio tends to a finite limit. The same property holds exactly in the case of a sufficiently small~$q>0$, with the finite limit being independent of~$q$ (see Appendix~\ref{corrections} for a proof). This allows us to evaluate exactly the maximal deviation from a circle for non-equatorial unstable spherical orbits with~$Q>>1$. Namely,
\beqa
\frac{\Delta Q}{Q} = \frac{a^2(E^2 - 1)\left.\cos^{2}{\theta}\right|_{\rm max}}{Q} = a^2(\epsilon^2 - q)\left.\cos^{2}{\theta}\right|_{\rm max} = \nonumber \\
= -\frac{1}{2}\left(\lambda_z^2 + 1 - a^2(\epsilon^2 - q) - \sqrt{\left[ 
\lambda_z^2 + 1 - a^2(\epsilon^2 - q)\right]^2 + 4a^2(\epsilon^2 - q)}\right) = \nonumber \\
= \frac{a^2}{(\lambda_z/\epsilon)^2 - a^2} + \mathcal{O}\left(\frac{1}{x^2}\right)\,, \quad x^2\to +\infty\,.
\eeqa
In this derivation we have used relations~(\ref{epsq}) and~(\ref{lambdaq}) from Appendix~{\ref{corrections}} to make sure that
\be
\lambda_z^2 + 1 - a^2(\epsilon^2 - q) =\mathcal{O}(x^2) > 0 \quad \mbox{as} \quad x^2\to +\infty\,.
\ee

Finally, we are in a position to find the maximal deviation. Using~(\ref{epsq0}) and~(\ref{lambdaq0}), we obtain for corotating orbits:
\beqa
\frac{\Delta Q_{\rm max^{+}}}{Q} &=& \lim\limits_{a\to 1^{-}}{\frac{a^2}{(\lambda_z/\epsilon)^2 - a^2}}=\frac 13 \quad \Rightarrow \nonumber \\
\nonumber \\
& \Rightarrow& \qquad \frac{\mathcal{L}_{\rm max} - \mathcal{L}_{\rm min}}{\mathcal{L}_{\rm max} + \mathcal{L}_{\rm min}} = \left(2-\sqrt{3}\right)^2 \approx 0.07\,.
\eeqa

For counterrotating orbits, the largest deviation is not necessarily given by the respective limit taken at~$y\to y_{-}$. Instead, orbits with~$\lambda_z<0$ and~$\lambda_z$ close to zero may give rise to larger deviations, at least for small~$q$. For the sake of completeness, let us consider both cases.

At~$y\to y_{-}$, provided that~$\epsilon\to+\infty$ and~$\lambda_z\to -\infty$, the answer reads:
\beqa
\frac{\Delta Q_{\rm max^{-}}}{Q} &=& \lim\limits_{a\to 1^{-}}{\frac{a^2}{(\lambda_z/\epsilon)^2 - a^2}}=\frac{1}{48} \quad \Rightarrow \nonumber \\
\nonumber \\ 
& \Rightarrow& \qquad \frac{\mathcal{L}_{\rm max} - \mathcal{L}_{\rm min}}{\mathcal{L}_{\rm max} + \mathcal{L}_{\rm min}} = \left(2-\sqrt{3}\right)^4 \approx 0.005\,.
\eeqa

For $\lambda_z<0, |\lambda_z|<<1$, it is convenient to write the answer in a parametric form. That is, a spherical orbit of radius~$1/y$ with~$\lambda_z=0$ has
\be
q = \frac{y(1+a^2y^2)(1 - 3 y + a^{2} y^{2} + a^{2} y^{3})}{(1+a^2y^2)^2 - 4a^2y^3}\geq 0\,, \qquad y\in[y_{-},y_{*}]\,,
\ee
where $y=y_{*}$ corresponds to~$q=0$ and satisfies $1 - 3 y + a^{2} y^{2} + a^{2} y^{3}=0$, cf.~(\ref{lambdaq0}). Notice also that the denominator is positive and the numerator has exactly one root for~$y\in(0,1)$, $a\neq 1$. The respective deviation is
\be
\frac{\Delta Q}{Q} = a^2(\epsilon^2 - q) = -\frac{a^2y^2\left[(1+a^2y^2)^2-4y\right]}{(1+a^2y^2)^2 - 4a^2y^3}\,. 
\ee
This deviation is maximal at~$q=0$ and~$a\to 1$, which implies~$y_{*} = \sqrt{2}-1$. Hence,  
\beqa
\frac{\Delta Q_{\rm max^{0}}}{Q} &=& -\left.\frac{a^2y^2\left[(1+a^2y^2)^2-4y\right]}{(1+a^2y^2)^2 - 4a^2y^3}\right|_{y=\sqrt{2}-1}=\frac{8\sqrt{2}-11}{7} \quad \Rightarrow \nonumber \\
&\Rightarrow& \quad \frac{\mathcal{L}_{\rm max} - \mathcal{L}_{\rm min}}{\mathcal{L}_{\rm max} + \mathcal{L}_{\rm min}}  \approx 0.01\,.
\eeqa

\begin{figure*}
\begin{center}
	\includegraphics[width=0.45\textwidth]{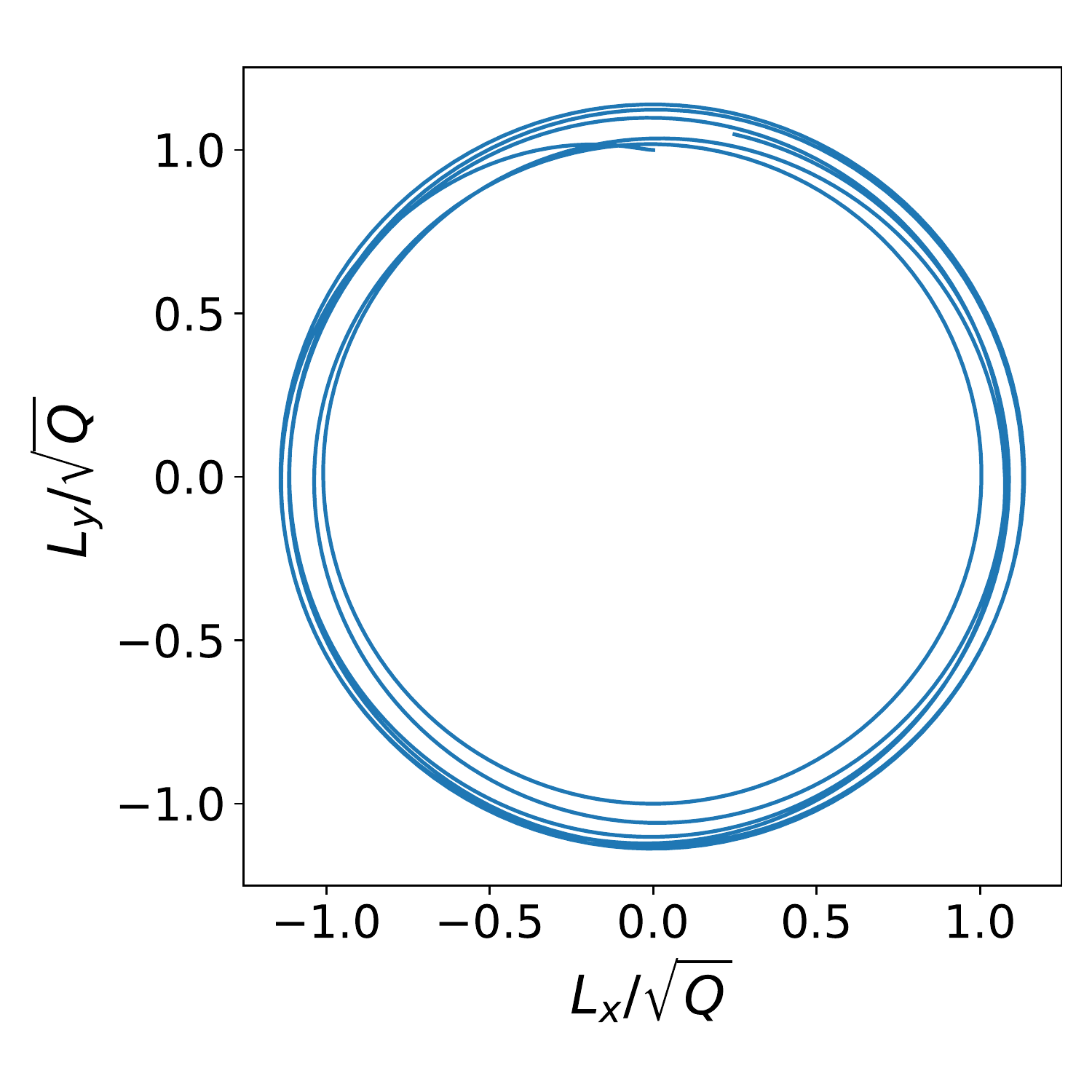}
	\includegraphics[width=0.45\textwidth]{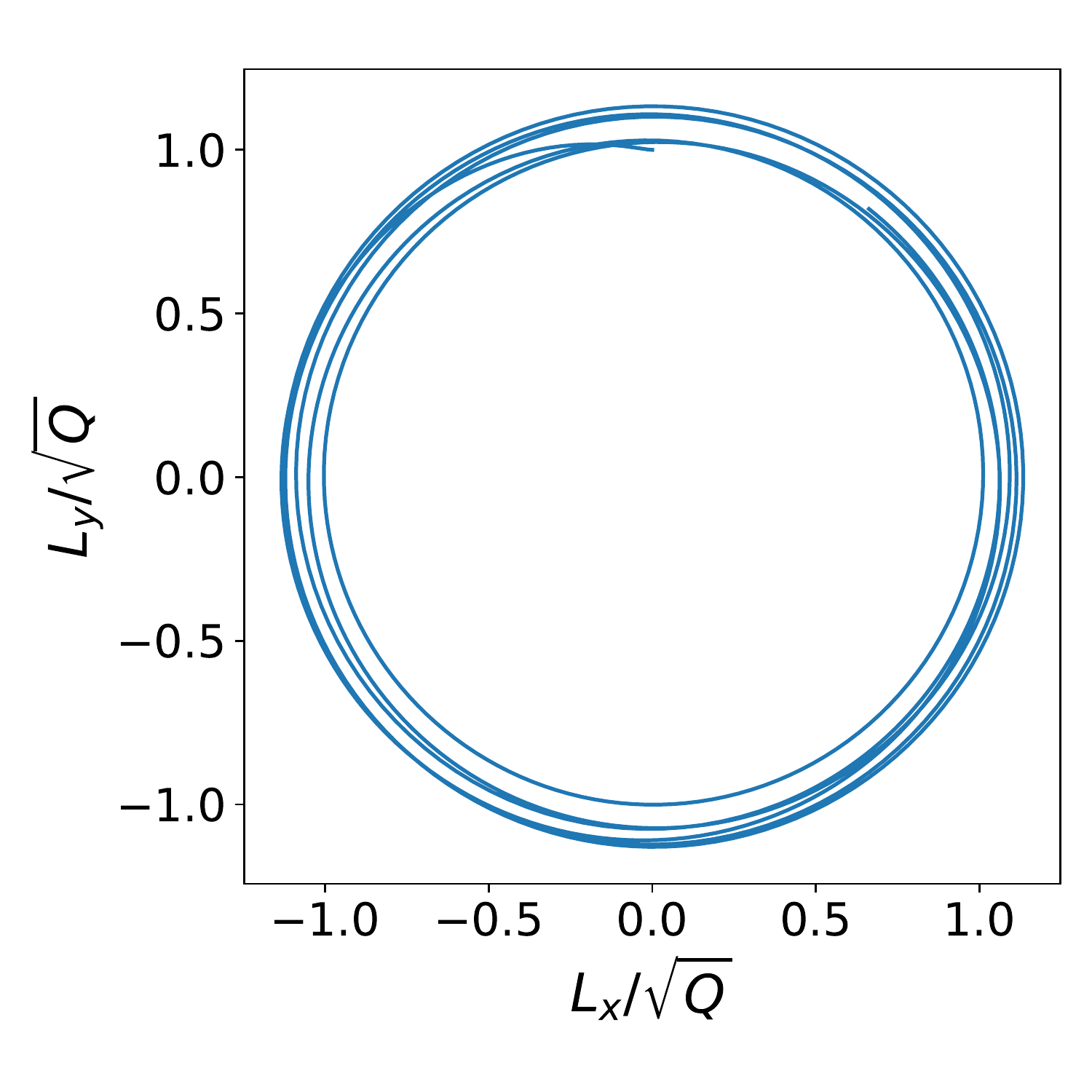}
\end{center}
	\caption{Maximally noncircular hodographs of a nearly equatorial orbit (\textit{left}) and an orbit with~$Q>>1$ (\textit{right}). Parameters of the nearly equatorial orbit are $E=100$, $L_z = 207.8128$, $Q=0.1728$. The second orbit has $E=19.5232$, $L_z = 40.5399$, $Q=50$. Both particles orbit a nearly extreme black hole with $a=0.999$ and start from~$r_0 = 10$, $\theta_0 = \pi/2$, and $\phi_0 = 0$.}
	\label{superrelativistic_unbounded}       
\end{figure*}

Concluding the case of unbounded orbits, their hodographs deviate from a circle by no more than~$\approx 7\%$. Figure~\ref{superrelativistic_unbounded} shows such close-to-extreme hodographs.


\section{Deviations from a circle}\label{nutation}

As we saw, the hodographs of the total angular momentum are nearly circular for both bounded and unbounded orbits that do not end up in the black hole. In any case the deviation from the circular shape does not exceed~$\approx 10\%$ and~$\approx 7\%$ for bounded and unbounded orbits, respectively. The circular shape implies precession while the deviations can be approximated by nutation (somewhat reminiscent of Ptolemaic epicycles). Interestingly, the more the deviation, the better nutation approximates it.

\begin{figure*}
\begin{center}
	\includegraphics[width=0.6\textwidth]{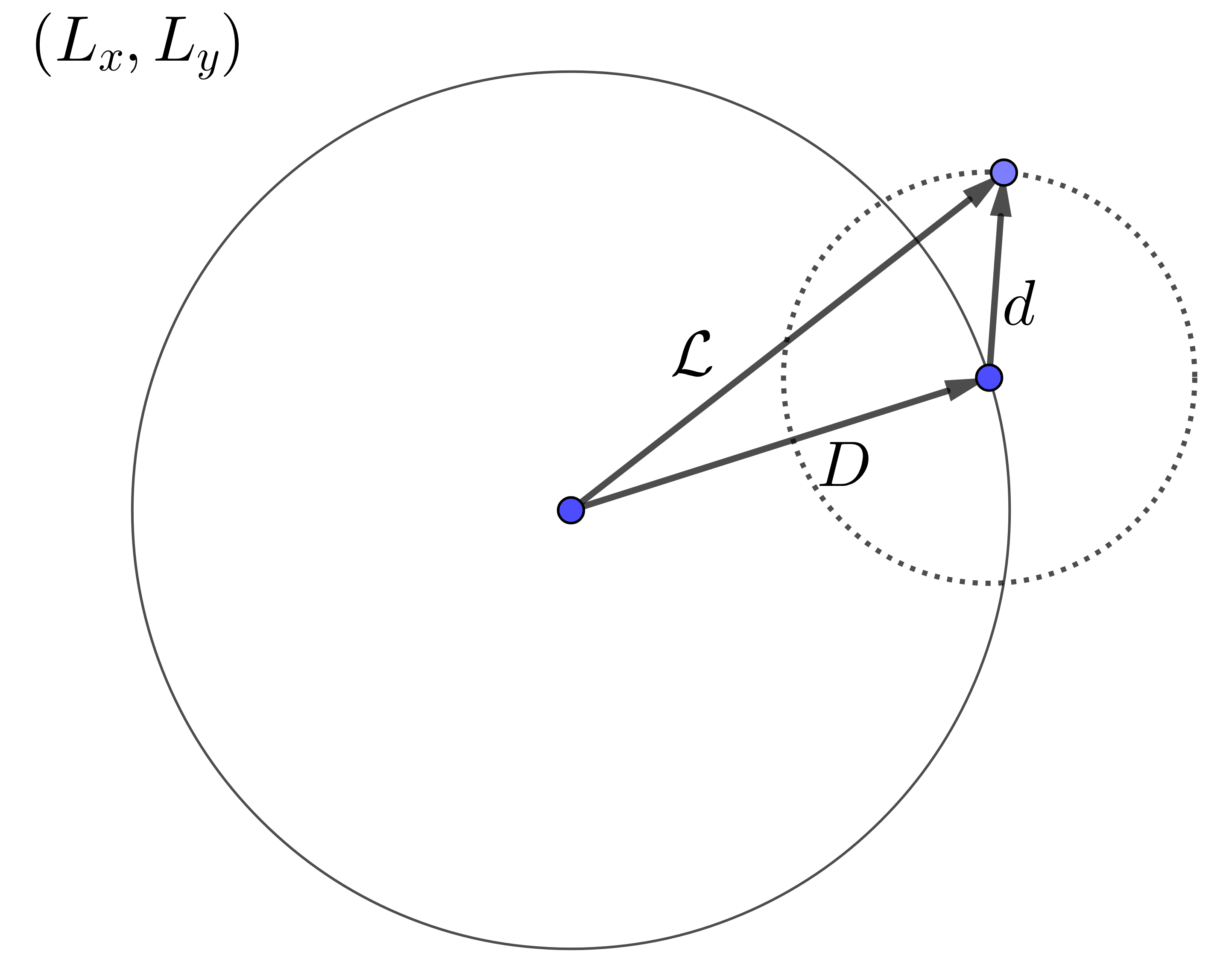}
\end{center}
	\caption{\textit{Left:} a hodograph deviating from a circle. \textit{Right:} Hodograph's motion represented as nutation along a circle of radius~$d$, the center of which undergoes precession along a circle of radius~$D$.}
	\label{fig:4}       
\end{figure*}

In the nutation approximation, the radius~$\mathcal{L}$ in the $(L_x,L_y)$-plane is represented as a sum of two constant-length vectors (see Figure~\ref{fig:4}). Precession vector~$\mathbf{D}$ rotates around the origin, whereas nutation vector~$\mathbf{d}$ rotates around the end of the precession vector.
\beqa
&\mathcal{L} = \mathbf{D}(\gamma) + \mathbf{d}(\gamma)& \\
&\Updownarrow& \nonumber \\
&L_x = D\cos(\Omega_{\rm p}\gamma + \psi_{\rm p}) + d\cos(\Omega_{\rm n}\gamma + \psi_{\rm n})\,,& \label{Lx_nut}\\
&L_y = D\sin(\Omega_{\rm p}\gamma + \psi_{\rm p}) + d\sin(\Omega_{\rm n}\gamma + \psi_{\rm n})\,,& \label{Ly_nut}
\eeqa
where $\Omega_{\rm p}$ and~$\Omega_{\rm n}$ are the precession and nutation frequencies, respectively; $\psi_{\rm p}$ and~$\psi_{\rm n}$ are the corresponding phase shifts, and $\gamma$ is a parameter along the trajectory of a particle such as $\dd\gamma = E\dd\tau/\rho^2$. Thus, the squared radius is given by
\be
\mathcal{L}^2 = D^2 + d^2 + 2D d\cos{(\gamma\Delta\Omega + \Delta\psi)}\,, \label{NutFunc}
\ee
where $\Delta\Omega\equiv\Omega_{\rm p} - \Omega_{\rm n}$ and $\Delta\psi\equiv\psi_{\rm p} - \psi_{\rm n}$.

Now, let us show that (\ref{NutFunc}) does provide an approximation for the actual~$\mathcal{L}$. Indeed, consider equation of motion~(\ref{eq:geodes2}) in the $\theta$-direction and rewrite it in terms of $\epsilon$, $\lambda_z$, $q$, $\mu\equiv\cos{\theta}$, and the new parameter~$\gamma$:
\be
\epsilon^2\left(\frac{\dd\mu}{\dd\gamma}\right)^2 + \mu^2\left(\lambda_z^2 + 1 - a^2(\epsilon^2 - q)\right) + \mu^4 a^2(\epsilon^2 - q) = 1\,.
\ee
In essence, this is the total energy of an anharmonic oscillator. Its degree of anharmonicity is given by the ratio of the third and second terms on the left-hand side:
\be
\frac{\mu_{\rm max}^2 a^2 (\epsilon^2 - q)}{\lambda_z^2 + 1 - a^2(\epsilon^2 - q)} = \frac{\Delta Q_{\rm max}/Q}{\lambda_z^2 + 1 - a^2(\epsilon^2 - q)}\,.
\ee
For both bounded and unbounded orbits with maximal deviation, the denominator is large while the numerator is on the order of one. Thus, the oscillator is approximately harmonic with the frequency 
\be
\omega_0 = \frac{\sqrt{\lambda_z^2 + 1 - a^2(\epsilon^2 - q)}}{\epsilon}\,.
\ee
Since the degree of anharmonicity is small, it can be taken into account by correcting the frequency. The answer is easy to find~\cite{mechanics}:
\be
\omega = \omega_0 + \frac 34\frac{\Delta Q_{\rm max}/Q}{\omega_0\epsilon^2}\,.
\ee

Assuming that a particle starts from the equatorial plane, we obtain \mbox{$\mu(\gamma) = \mu_{\rm max}\sin{\omega\gamma}$}, which results in
\be
\label{nut_approx}
\frac{\mathcal{L}^2}{Q} = 1 + \frac 12\frac{\Delta Q_{\rm max}}{Q}  - \frac 12\frac{\Delta Q_{\rm max}}{Q}\cos{2\omega\gamma}\,.
\ee
That is, the hodograph's motion is a sum of precession and nutation with $\Delta\Omega = 2\omega$, cf.~(\ref{NutFunc}). The relative phase shift~$\Delta\psi$ is fixed to~$0$ by choosing the equatorial plane as a starting point. 

\begin{figure*}[h]
	\includegraphics[width=0.5\textwidth]{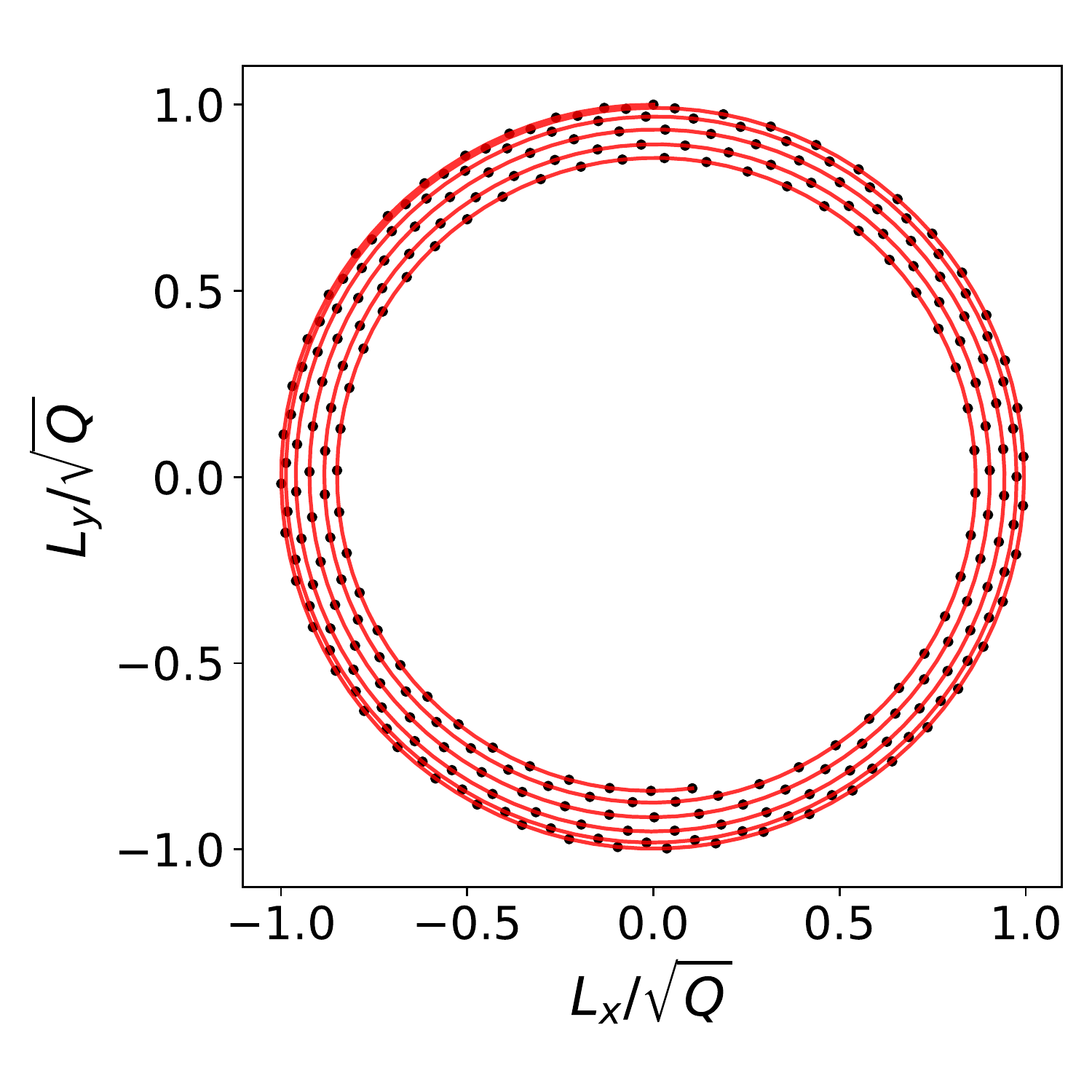}
	\includegraphics[width=0.5\textwidth]{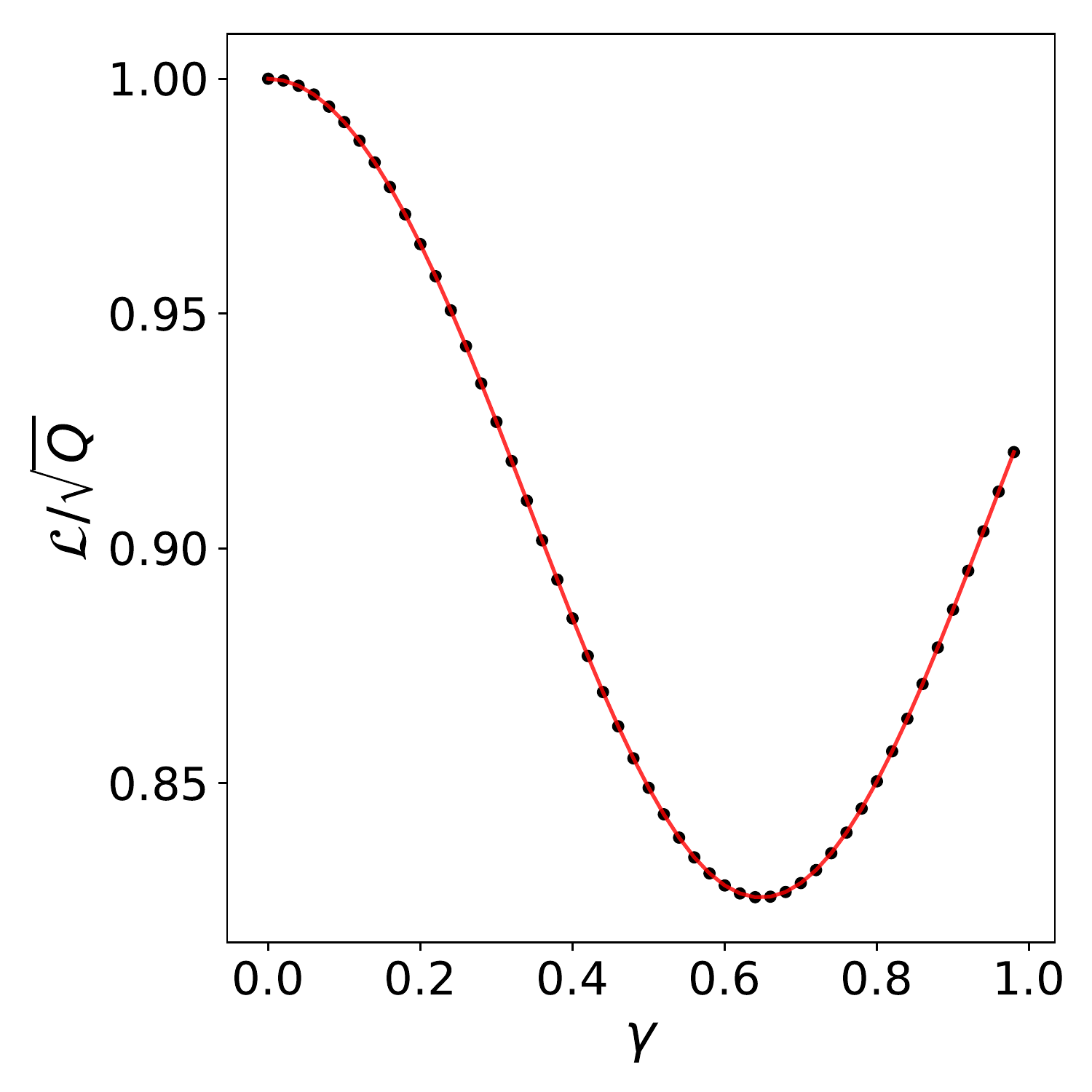}
	\caption{\textit{Left:} A hodograph of a bounded orbit approximated by a sum of precession and nutation. \textit{Right:} The squared radius in $(L_x,L_y)$-plane as a function of integration parameter~$\gamma$. Dots depict the exact curves obtained through numeric simulation by the classical fourth-order Runge--Kutta method with absolute numeric error estimated to be $<10^{-7}$, according to Runge's rule~\cite{Hairer}. Red line shows the approximation given by eq.~(\ref{nut_approx}). Parameters of the orbit are those of Figure~\ref{superrelativistic_bounded}.  }
	\label{bounded_nut}       
\end{figure*}

\begin{figure*}[h]
	\includegraphics[width=0.5\textwidth]{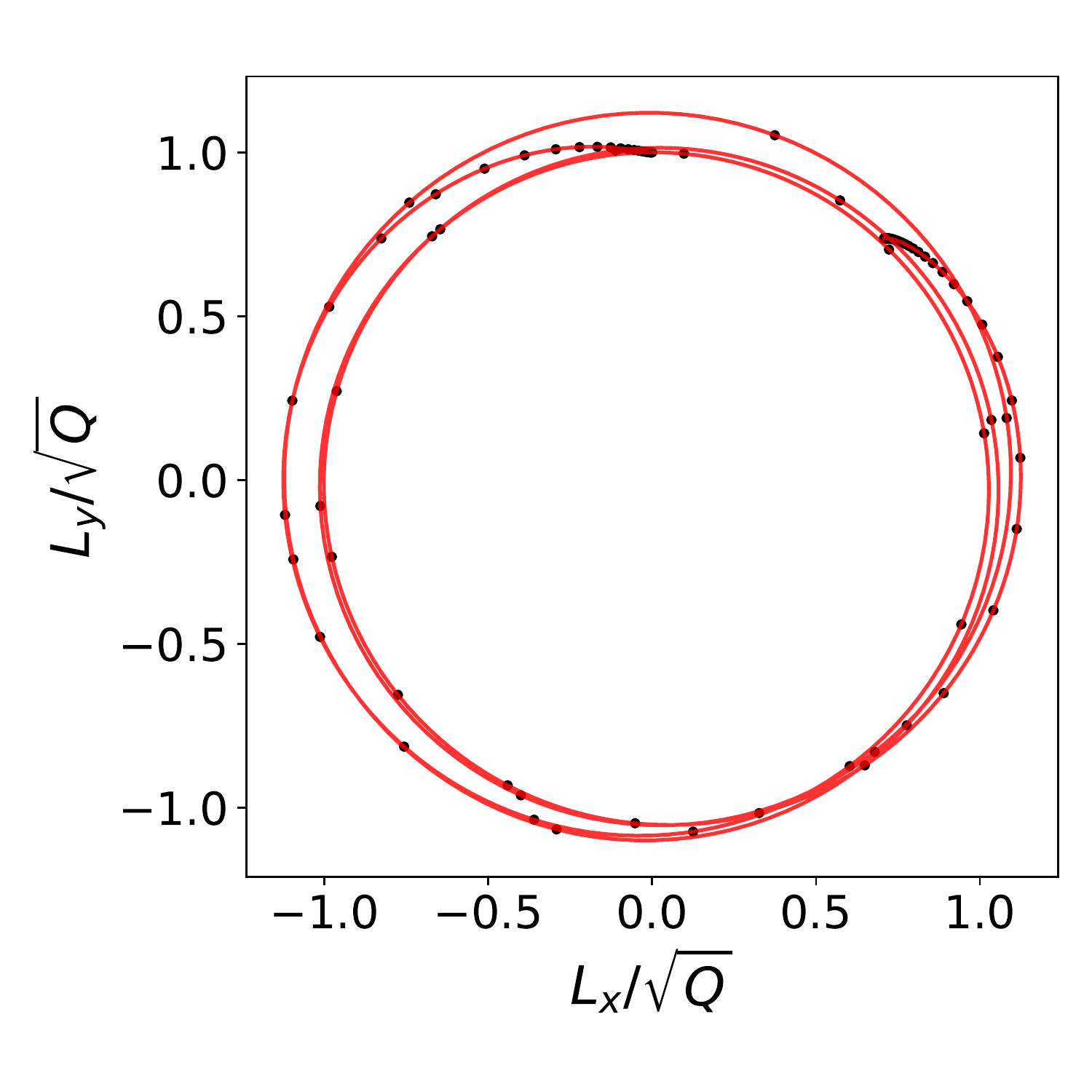}
	\includegraphics[width=0.5\textwidth]{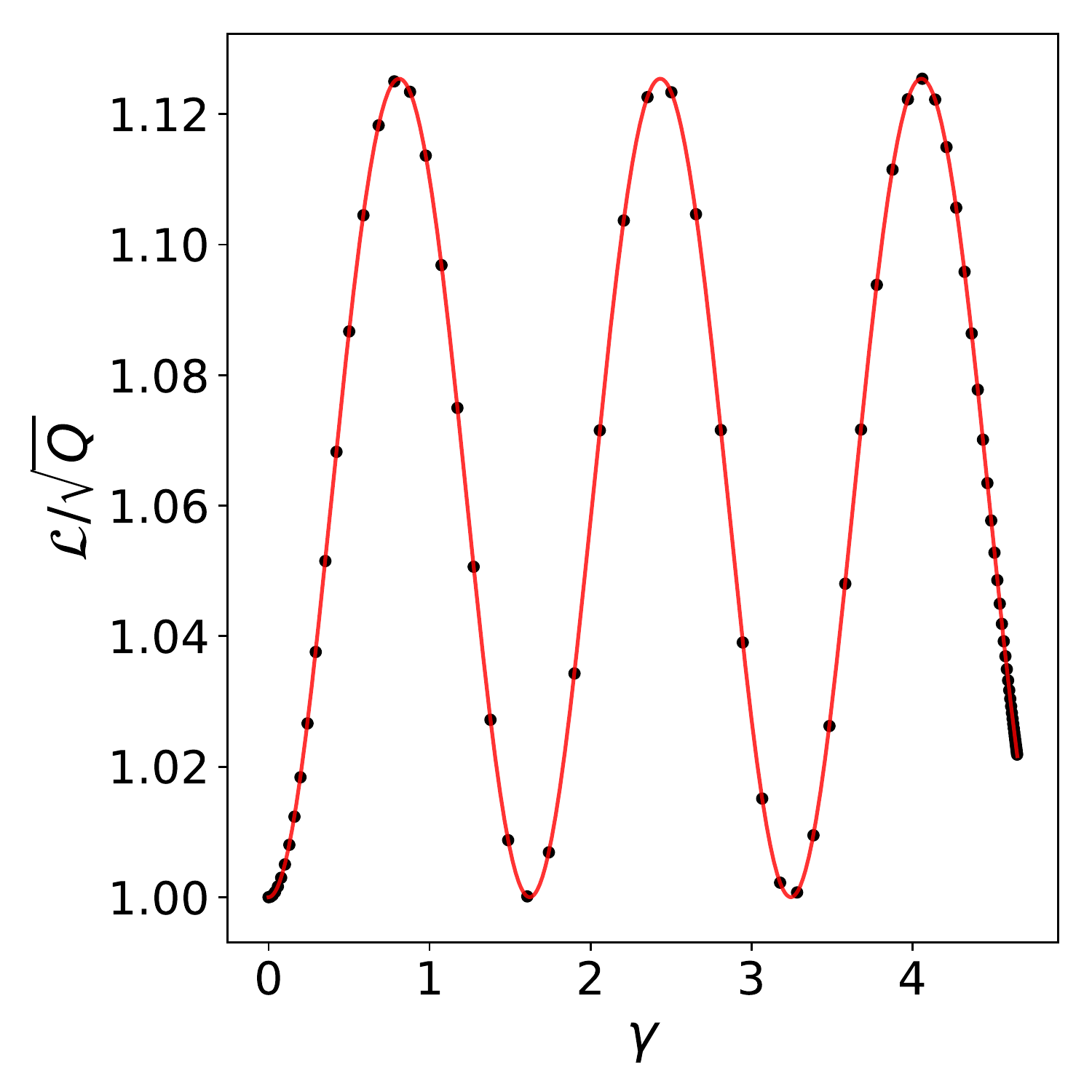}
	\caption{\textit{Left:} A hodograph of an unbounded orbit approximated by a sum of precession and nutation. \textit{Right:} The squared radius in $(L_x,L_y)$-plane as a function of integration parameter~$\gamma$. Dots depict the exact curves obtained through numeric simulation by a variable step fourth-order Runge--Kutta method~\cite{Hairer} with absolute numeric error  $<10^{-10}$. Red line shows the approximation given by eq.~(\ref{nut_approx}). Parameters of the orbit are those of the second orbit of Figure~\ref{superrelativistic_unbounded} except for energy~$E=19$.  }
	\label{unbounded_nut}       
\end{figure*}

Therefore, either the deviation is small and the hodograph is approximated by a circle or the deviation is large and nutation almost perfectly accounts for a noncircular shape. Figures~\ref{bounded_nut} and~\ref{unbounded_nut} show actual hodographs and their approximations by a sum of precession and nutation for bounded and unbounded orbits, respectively. The r.m.s. of the approximation's relative error is~$\lesssim 10^{-4}$.


\section{The order of magnitude for the nutation}\label{observe_nut}

The Lense--Thirring precession frequency~$\Omega_{\rm p}\sim GJ/(c^2r^3)$, where $J$ is the angular momentum of the gravitating center and $r$ is the characteristic size of the orbit around that center, e.g.~\cite[\S~40.7]{MTW}. For Earth, the magnitude of the precession $\sim 100$~milliarcseconds per year, which was confirmed through measurements of the effect on Earth-orbiting satellites~\cite{Ciufolini_2004,Everitt_2011}.

To estimate the nutation frequency~$\Omega_{\rm n}$, consider the angular velocity of vector~$\mathcal{L}$ in the $(L_x,L_y)$-plane. Let $f\equiv\arctan{\frac{L_y}{L_x}}$. Then, by virtue of~(\ref{Lx_nut}) and~(\ref{Ly_nut}),
\beqa
\mathcal{L}^2\frac{df}{d\gamma} &=& L_{x}\frac{dL_y}{d\gamma} - L_{y}\frac{dL_x}{d\gamma}\, \nonumber \\
&=& D^2\Omega_{\rm p} + d^2\Omega_{\rm n} + Dd(\Omega_{\rm p} + \Omega_{\rm n})\cos{(\gamma\Delta\Omega + \Delta\psi)}\,.
\eeqa
Now, when $\mathcal{L}^2$ is maximal or minimal (the upper or lower sign in what follows),
\beqa
\mathcal{L}^2_{\rm m} &=& D^2 + d^2 \pm 2Dd\,, \\
\mathcal{L}^2_{\rm m}\left.\frac{df}{d\gamma}\right|_{\rm m} &=& D^2\Omega_{\rm p} + d^2\Omega_{\rm n} \pm Dd(\Omega_{\rm p} + \Omega_{\rm n})\,,
\eeqa
whence
\beqa
\Omega_{\rm p} &=& \left.\frac{df}{d\gamma}\right|_{\rm m} \pm \frac{d\Delta\Omega}{D \pm d}\,, \\
\Omega_{\rm n} &=& \left.\frac{df}{d\gamma}\right|_{\rm m} - \frac{D\Delta\Omega}{D \pm d}\,. \\
\eeqa

On the other hand, a direct differentiation of~(\ref{eq:Lx}) and~(\ref{eq:Ly}) yields
\be
\mathcal{L}^2\frac{df}{d\gamma} = \frac{\rho^2 p^\phi}{E}\cdot(p_\theta^2 + p_\phi^2\cot^{2}{\theta}) - \frac{p_\phi}{E}\left(\frac{p_\theta^2}{\sin^{2}{\theta}} + \rho^2\dot{p}_{\theta}\cot{\theta}\right),
\ee
which, after appropriate substitutions from~(\ref{eq:Q}), (\ref{eq:geodes2}), and~(\ref{eq:geodes3}), results in
\be
\frac{df}{d\gamma} = \frac{a(2rE-aL_z)}{E\Delta} + \frac{L_z}{E}\frac{a^2(E^2 - 1)\cos^{2}{\theta}}{Q + a^2(E^2 - 1)\cos^{2}{\theta}}\,.
\ee

At the starting point~$\theta=\pi/2$, derivative~$df/d\gamma$ is maximal (unbounded orbits) or minimal (bounded ones). In either case,
\beqa
\Omega_{\rm p} &=& \frac{a(2rE-aL_z)}{E\Delta} - \omega\left(\sqrt{1 + \frac{\Delta Q_{\rm max}}{Q}} - 1\right)\,, \\
\Omega_{\rm n} &=& \frac{a(2rE-aL_z)}{E\Delta} - \omega\left(\sqrt{1 + \frac{\Delta Q_{\rm max}}{Q}} + 1\right)\,.
\eeqa
Note that these relations give an order-of-magnitude estimate of the frequencies, where $r$ is a characteristic size of the orbit. For the unbounded orbits, the frequencies vary significantly as the particle approaches the black hole from infinity. This is because the orbital plane is approximately the same when the particle is far from the black hole. Figure~\ref{unbounded_nut} illustrates the fact: the start and the end of the trajectory manifest themselves by portions of the graphs with many integration points (in black).

In order to estimate the frequencies for Earth-orbiting satellites, we retain the first nonvanishing terms with respect to~$a$ and express the frequencies with respect to proper time~$\tau$:
\beqa
\Omega_{\rm p}^{E} &\approx& \frac{2a}{r^3} \,, \label{freq_precession}\\
\Omega_{\rm n}^{E} &\approx& - \frac{2\omega_0}{r^2} \approx - \frac{2L}{r^2}\,,
\eeqa
where we have used the fact that the size of the orbit~$r>>1$, total energy $E\approx 1$, and $L_z^2 + Q$ equals the squared total angular momentum~$L^2$ up to terms $\propto a^2$. The precession frequency is obviously the Lense--Thirring frequency in a dimensionless form.

As for the nutation frequency, it is negative and its (dimensional) magnitude is
\be
|\Omega_{\rm n}^{E}| = \frac{2v}{r} = \frac{4\pi}{T}\,,
\ee
where $T$ is the orbital period of a satellite on a circular orbit. That is, this frequency is twice the orbital frequency, and the nutation occurs in the direction opposite to that of the precession. For orbits with moderate inclinations~$i$, the nutation effect would lead to slight oscillations of the orbital plane by an angle
\beqa
\Delta\theta &=& \frac{|\Delta Q_{\rm max}|\cos{i}}{2L\sqrt{Q}} = \frac{a^2|E^2 - 1|\sin^{2}{i}\cos{i}}{2L^2\sin{i}} \nonumber \\
 &=& \frac{a^2\sin{2i}}{4L^4}\,,
\eeqa
where we have substituted $|E^2 - 1| = 1/L^2$ for circular orbits (in the dimensionless units). Or, in the dimensional form:
\be
\Delta\theta \sim \left(\frac{GJ}{c}\right)^2\frac{1}{l^4} \sim \left(\frac{2\pi R_0}{cT_0}\right)^2\left(\frac{R}{r}\right)^2\,,
\ee
where $l$ is the specific orbital angular momentum (per mass of the orbiting body) and $R_0, T_0$ are the radius and rotation period of the gravitating center, respectively.

The linear displacement of the orbiting body due to the nutation is in turn
\beqa
\Delta r_{\perp}^{\rm{Earth}} &\sim& 10^{-6}\frac{R_{\rm E}}{r}\;\mbox{[meters]}\,, \\
\Delta r_{\perp}^{\rm{Jupiter}} &\sim& 10^{-2}\frac{R_{\rm J}}{r}\;\mbox{[meters]}\,, \\
\Delta r_{\perp}^{\rm{WD}} &\sim& 10^{-2}\frac{R_{\rm WD}}{r}\;\mbox{[meters]}\,, \\
\Delta r_{\perp}^{\rm{NS}} &\sim& 100\frac{R_{\rm NS}}{r}\;\mbox{[meters]}\,, \label{amp_NS}
\eeqa
where NS and WD stand for ``neutron star'' and ``white dwarf'', respectively. For the latter, we took white dwarf~G29-38 as a representative with $R_0 = 0.01R_{\astrosun}$ and $T_0 = 0.014$~days~\cite{Reach_2009,Koester_1998} and we adopted that, first, a solar-mass neutron star with radius of $10$~km rotates at the maximum possible speed ($T_0 \sim 1$~ms) and, second, the dimensionless moments of inertia of Earth, Jupiter, the neutron star, and the white dwarf are $0.35$.

\section{Conclusion}\label{conclusion}

We have studied the evolution of the orbital angular momentum of a massive particle following a geodesic path in the Kerr space--time. In doing so, we restricted ourselves to orbits, both bounded and unbounded, that do not end up in the black hole. 

We have found that the end of the angular momentum approximately describes a circle, such that it deviates from the circle by no more than~$\approx 10\%$ and~$\approx 7\%$ for bounded and unbounded orbits, respectively. Curiously, for corotating orbits the relative maximal deviation takes a universal value independent of the type of orbit, $|\Delta Q|_{\rm max^{+}}/Q = 1/3$. Second, we have demonstrated that nutation (precession around a precessing axis) accounts for this deviation within~$0.01\%$ for the orbits with maximal deviation. These results imply that the Lense--Thirring precession of the orbital angular momentum, originally found in the weak-field limit, continues to be a valid description in the general case of (almost) arbitrary exact orbits. Also, introducing nutation makes this description highly accurate.

We have also estimated the nutation frequency and magnitude in the limit $a/r<<1$, where $r$ is a characteristic size of the orbit, equations~(\ref{freq_precession})--(\ref{amp_NS}). It turns out that the frequency is twice the orbital frequency in this limit. As for the magnitudes, the corresponding linear displacement for a satellite orbiting the Earth is $\sim 10^{-6}$~m, which eliminates any prospect of detecting the effect in the near future. For Jupiter and a white dwarf, this displacement is $\sim 10^{-2}$~m, which is, in principle, on the order of the available precision of a few millimeters~\cite{Ciufolini_2009}. However, the practical prospects are vague, given that there is no Jupiter's gravity model similar to GRACE. For a neutron star, however, the linear displacement is considerable ($\sim 100$~m), which could be detected through the Doppler shift $\sim \Delta r_{\perp}^{\rm{NS}}/T_0$. The latter is comparable to the expected precision of third-generation spectohraphs ($\lesssim 10$~cm/s for ESPRESSO~\cite{Pepe_2014}), provided that the test body orbits the neutron star at a distance $\sim 500$ star radii. We leave a detailed study of radial velocity profiles for this case to future research.

Although this paper deals with the orbital Lense--Thirring precession, the effect of nutation as found in this work is somewhat reminiscent of the relativistic nutation of a gyroscope. Recall that the angular momentum of the Earth--Moon system plays the role of a gyroscope moving in the gravitational field of the Sun, which undergoes the geodetic~\cite{de_Sitter_1916,Bertotti_1987,Shapiro_1988} as well as a gravitomagnetic precession (see, for example,~\cite{Mashhoon_1986}). In the motion of the gyroscope, there are also the effects of relativistic nutation which are long- and short-period. The former combines with the precession at short timescales (see~\cite{Mashhoon} and references therein), whereas the latter has the frequency twice that of the orbital motion and is effectively averaged out~\cite[p.~550]{Mashhoon_1991}. 

The amplitude of these kinds of nutation is linear in~$a$. Another kind of nutation with an amplitude proportional to the quadrupole moment of the central gravitating body (and, hence, to~$a^2$) may result from a contribution to the tidal matrix which is periodic with a frequency twice that of the orbital motion~\cite[Appendix~C]{Mashhoon_1991}. Thus, at least the frequency of certain kinds of the relativistic nutation of a gyroscope is the same as the nutation frequency in the orbital Lense--Thirring effect (see Sec.~\ref{observe_nut}). However, it remains unclear whether there is a deeper connection between the nutation of a gyroscope and the orbital nutation\,\footnote{The precession of a gyroscope on essentially relativistic bounded orbits around a Kerr black hole was studied in~\cite{Bini_2016}. However, the explicit expressions are quite cumbersome (even when restricted to the case of equatorial plane), and the properties of the nutation in this fully relativistic case are not evident.}.


Finally, as a by-product, we have also derived the parameters of unstable spherical timelike orbits for arbitrary rotation parameter~$a$ and Carter's constant~$Q=1/q$ in an alternative form, i.e. as a function of~$a$, $q$, and the radius of the unstable circular orbit, see~(\ref{rootq}) and~(\ref{xq})--(\ref{lambdaq}). These relations provide an analytical way to separate unbounded orbits that will end in the black from those that will not.  Previously, such relations were either restricted to extremely rotating black holes~\cite{Wilkins_1972,Johnston_1974} or given as a function of the radius and energy~$E$ of the particle~\cite{chandrasekhar2}. The alternative form of those relations has allowed us to notice a remarkable fact. Namely, if $q = 1/Q \leq ay^{5/2}_{-}$, the orbit of a particle abruptly becomes confined to a plane as~$r\to r_{\rm ph}^{\pm}$, even though $Q$ is large. This is not so if $a=1$ or $q > ay^{5/2}_{-}$ at either $y_{-}$ or $y_{+}$. This latter case requires further investigation.
%

\section*{Acknowledgements}
The analytic part of this research was carried out by VNS and supported by a grant of the Russian Science Foundation (project no.~17-71-10260). The numeric simulations were carried out by SKh.

We are grateful to V.S. Beskin and Yu.Yu. Kovalev for providing a work en\-viron\-ment which made completion of this Paper possible. We thank S.V. Repin for useful comments regarding the numerical simulation of geodesics and a presentation of this work before a committee. Last but not least, we thank an anonymous reviewer for pointing out (among other constructive comments) the analogy between the orbital Lense--Thirring precession and the precession of a gyroscope.

Symbolic computations were partially performed with SymPy~\cite{sympy}. We also made use of IPython \cite{PER-GRA:2007}, SciPy \cite{jones_scipy_2001},  Matplotlib \cite{Hunter:2007}, NumPy \cite{van2011numpy}, and this preprint was typeset in \texttt{arxiv-style}~\cite{arxiv_sty}.

\appendix

\section{Numeric solution of geodesic equations}\label{numeric_geodesics}

The geodesic equations can be recast into a form that is more convenient for numeric simulations~\cite{Sharp_1981,Zakharov_1994,zakharov_repin}. Although the number of equations increases, now there is no need to track signs of the square roots present in the initial form. The equations to be solved are 
\beqa
\frac{dr}{d\gamma}&=&r_1 \label{first}\\
\frac{d r_1}{d\gamma}&=& \frac{1}{2E}\frac{\dd R(r)}{\dd r}= \nonumber \\
&= & 2r^3 + r\left(a^2 - \xi^2 - \eta\right)+ \left(a-\xi\right)^2 + \eta - \frac{r}{E^2}\left(2r^2 - 3r + a^2\right) \\
\frac{d\theta}{d\gamma}&=& \theta_1 \\
\frac{d\theta_1}{d\gamma}&=&\frac{1}{2E}\frac{\dd \Theta(\theta)}{\dd\theta}=\cos\theta\left[a^2(E^{-2}-1)\sin\theta+\frac{\xi^2}{\sin^3\theta}\right]\\
\frac{d\phi}{d\gamma}&=& \frac{\xi}{\sin^2\theta} - a +\frac{a(r^2+a^2-\xi a)}{r^2-2r+a^2}\,, \label{last}
\eeqa
where $d\tau/d\gamma=\rho^2/E$, $\xi = L_z/E$, and $\eta=Q/E^2$. Also, the following relations hold between the functions:
\beqa
r_1^2&=&  r^4 + \left(a^2 - \xi^2 - \eta\right)r^2 + \nonumber \\
&+& 2\left[(a-\xi)^2 + \eta\right]r - a^2\eta - \frac{r^2}{E^2}\left( r^2-2r+a^2\right)\,, \label{eps1_r}\\
\theta_1^2&= &  \eta - \cos^{2}{\theta}\left[a^2(E^{-2}-1) + \frac{\xi^2}{\sin^{2}{\theta}}\right]\,. \label{eps2_r}
\eeqa

This set of differential equations was numerically solved with either the classical or variable step~\cite{Hairer} fourth-order Runge--Kutta method. That numeric solution was then used to draw the hodographs in Figures~\ref{superrelativistic_bounded}, \ref{superrelativistic_unbounded}, \ref{bounded_nut}, and~\ref{unbounded_nut}. For convenience and reproducibility, we summarize the parameters for those hodographs in Table~\ref{init_conds}. Together with~$r_1$ and~$\theta_1$ evaluated through relations~(\ref{eps1_r}) and~(\ref{eps2_r}), they yield initial conditions for equations of motion~(\ref{first})--(\ref{last}). Also, we always choose~$r_1<0$ and~$\theta_1>0$ at the start.

\begin{table}
\centering\caption{Initial conditions for orbits with noncircular hodographs.\label{init_conds}}
\begin{tabular}{@{}*{8}{l}}
\hline\noalign{\smallskip}
Figure & $a$ & $E$ & $L_z$ & $Q$ & $r_0$ & $\theta_0$ & $\phi_0$ \\
\noalign{\smallskip}\hline\noalign{\smallskip} 
\ref{superrelativistic_bounded} & 1 & 0.59     &  1.18026 & 0.00437 & 1.03 & $\pi/2$ & 0 \\
\ref{superrelativistic_unbounded} (\textit{left})       & 0.999 & 100      &  207.8128   & 0.1728     & 10 & $\pi/2$ & 0 \\
\ref{superrelativistic_unbounded} (\textit{right})       & 0.999 & 19.5232      &  40.5399    & 50      & 10 & $\pi/2$ & 0 \\
\ref{bounded_nut}        & 1 & 0.59 &  1.18026    & 0.00437 & 1.03    & $\pi/2$ & 0 \\
\ref{unbounded_nut}        & 0.999 & 19 & 40.5399    & 50 & 10 & $\pi/2$ & 0 \\
\hline\noalign{\smallskip}
\end{tabular}
\end{table}

\section{Behavior of functions $E(y,a)$ and $L_z(y,a)$ \\ (the parameters of stable circular equatorial orbits) }\label{EL_minimum}

Consider functions $E(y,a)$ and $L_z(y,a)$, which are given by~(\ref{eq:Ecirc}) and~(\ref{eq:Lcirc}), in the range~$y\in (0,1/r_{\rm ph}^{\pm})$. Their derivatives are
\beqa
\frac{\partial E}{\partial y} &=& -\frac{1 - 6y \pm 8ay^{3/2} - 3a^2y^2}{2\left(1 - 3y \pm 2ay^{3/2}\right)^{3/2}}\,, \\
\frac{\partial L_z}{\partial y} &=& \mp\frac{(1 - 6y \pm 8ay^{3/2} - 3a^2y^2)(1\pm ay^{3/2})}{2\left[y\left(1 - 3y \pm 2ay^{3/2}\right)\right]^{3/2}}\,.
\eeqa
They vanish simultaneously at $y = 1/r_{\rm isco}^{\pm}$, which satisfies~\cite{Bardeen_1972} 
\be
1 - 6y \pm 8ay^{3/2} - 3a^2y^2 = 0\,.
\ee
Besides, $E\to 1$ and~$L_z\to \pm\infty$ as~$y\to 0$, and~$E\to +\infty$ and~$L_z\to \pm\infty$ as $y\to y_{\pm} = 1/r_{\rm ph}^{\pm}$. Therefore, at $y = 1/r_{\rm isco}^{\pm}$, $E(y,a)$ must have a minimum, and~$L_z(y,a)$ must have a minimum (maximum) at~$y = 1/r_{\rm isco}^{+}$ ($y = 1/r_{\rm isco}^{-}$). In other words, both~$E(y,a)$ and~$L_z^{2}(y,a)$ have a minimum at~$y = 1/r_{\rm isco}^{\pm}$.

\section{Parameters of unstable spherical orbits with~$E>1$ and arbitrary $Q$}\label{corrections}

Consider conditions~(\ref{RQ}) and~(\ref{RQdiff}) without omitting terms~$\propto q\equiv 1/Q$:
\beqa
&{}&\epsilon^{2} - 2 \epsilon x\cdot a y^{2}  -  x^{2}y^{2}(1  - 2 y) - y^{2}\Delta_y- q\Delta_y = 0\,, \label{RQq}\\
&{}& - 2 \epsilon x \cdot a y  - x^{2}y(1- 3 y)  
- y (1 - 3 y + 2 a^{2} y^{2}) + q(1-a^2 y) = 0 \,,
\eeqa
where $\Delta_y\equiv a^2y^2 - 2y + 1 > 0$.

Solving the second of the equations for~$\epsilon$ and subtituting the solution to the first equation leads to a quadratic equation with respect to~$x^2$:
\be
Ax^4 + 2Kx^2 + C =0\,, \label{quadratic}
\ee
where
\beqa
A &\equiv& -(4 a^{2} y^{3} - 9 y^{2} + 6 y - 1)<0\,, \qquad 1/r_{\rm ph}^{-} < y < 1/r_{\rm ph}^{+}\,, \\
K &\equiv& 2a^2y^2\Delta_y + A - q\left(a^2 + a^2y - \frac{3y-1}{y}\right)\,, \\
C &\equiv& \left(2a^2y^2 - 3y + 1 + \frac{q(a^2y - 1)}{y}\right)^2\geq 0\,.
\eeqa

Let us now study how solutions to the quadratic equation behave at $y_{\pm}=1/r_{\rm ph}^{\pm}$. First, note that
\be
\lim\limits_{y\to y_{\pm}}{A(y)} =\lim\limits_{y\to y_{\pm}}{[(3y-1)^2 - 4a^2y^3]}=0\,.
\ee
Since $C > 0$ and $A\to 0^{-}$ as $y\to y_{\pm}$, the roots of~(\ref{quadratic}) have opposite signs. One of them tends to infinity and the other tends to a finite number. Their specific signs depend on the sign of~$K$. If $K>0$, the infinite root is positive and the finite root is negative. Otherwise, the infinite root is negative and the finite root is positive. If $K=0$, both roots tend to infinities of opposite signs.

If $q=0$ or, by continuity, $q<<1$, then $\lim\limits_{y\to y_{\pm}}{K}>0$, and at both $y_{-}$ and $y_{+}$ there is a positive infinite root. This root is the one that leads to solution~(\ref{xq0})--(\ref{lambdaq0}) obtained under assumption~$q<<1$.

For arbitrary $q>0$, it proves helpful to actually solve equation~(\ref{quadratic}). A positive root is found from relation
\beqa
-x^2 A = (2a^2y^2 - 3y + 1)^2 &-& q\left(a^2y+a^2 - \frac{3y-1}{y}\right) + \nonumber \\
 &+& 2a\Delta_y\left[\frac{\sqrt{\left(2q + y(3y-1)\right)^2 - y^2 A}}{2\sqrt{y}}-ay^2\right]\,. \label{rootq}
\eeqa

At $y_{+}$, which satisfies $3y-1 = 2ay^{3/2}>0$, the coefficient that multiplies~$q$ is reduced as follows:
\beqa
-a^2y-a^2 + \frac{3y-1}{y} + \frac{2a\Delta_y}{y^{1/2}} = -a^2y-a^2 - 2ay^{1/2} + 2a^3y^{3/2} + \frac{2a}{y^{1/2}} =  \nonumber \\
= -a^2y-a^2 - 2ay^{1/2} + a^2(3y-1) + \frac{2a}{y^{1/2}} = \frac{2a(1-y)(1-ay^{1/2})}{y^{1/2}} =\nonumber \\
=  \frac{ay(1 + y - 2ay^{1/2})}{y^{3/2}} =\frac{a(y(1+y)-3y+1)}{y^{3/2}} = \frac{a(1-y)^2}{y^{3/2}}\,.
\eeqa
Hence, as soon as $0<a<1$, 
\be
x^2 = -\frac{(2a^2y^2 - 3y + 1)^2 + \frac{a(1-y)^2}{y^{3/2}}q}{A} \to +\infty \quad \mbox{as} \quad A\to 0^{-}\,.
\ee

At $y_{-}$, which satisfies $3y-1 = -2ay^{3/2}<0$, the result depends on the magnitude of~$q$. If $q\geq -y(3y-1)/2 = ay^{5/2}$, a similar calculation leads to a right-hand side of~(\ref{rootq}) that vanishes as $A\to 0^{-}$. Thus, in this case the root tends to a finite positive value. Otherwise, the root tends to an infinite positive value. 

The function~$a[y_{-}(a)]^{5/2}$ grows in the range~$0\leq a\leq 1$ reaching its maximal value $1/32$ at $a=1$. Consequently, for a given~$a\in(0,1)$, there is a critical value of~$q$ that is below~$1/32$, which separates the cases of bounded and unbounded growth of~$x^2$ in a neighborhood of~$y_{-}$. Also, if $q$ is sufficiently small, the growth is always unbounded, which corresponds to the limiting case represented by~(\ref{xq0})--(\ref{lambdaq0}).

Another feature of~$x^2$ is that, for sufficiently small~$q$, it vanishes at a point~$y_0\in(y_{-},y_{+})$ where it also reaches a minimum. Indeed, for~$q=0$ it is trivial to verify this. If~$q<<1, q\neq 0$, the minimum value must simultaneously satisfy equations
\beqa
Ax^4 + 2Kx^2 + C =0\,, \\
x^4\frac{\partial A}{\partial y} + 2x^2\frac{\partial K}{\partial y} + \frac{\partial C}{\partial y} =0\,.
\eeqa
Since $C$ and $\partial C/\partial y$ have a common root, $x^2=0$ satisfies this last set of equations at that common root and, by continuity, is a minimum of the non-negative function~$x^2$. The case of~$q$ that is large enough for the common root to become complex is beyond the scope of this article.

To ensure that~$\epsilon>0$, the following signs must be chosen for~$x$:
\be
\label{xq}
x = \left\{
\begin{array}{ll}
\sqrt{x^2}\,, & \quad y_0\leq y< y_{+}\,, \\
-\sqrt{x^2}\,, & \quad y_{-}<y<y_0\,. 
\end{array}
\right.
\ee
Finally, the~$x$ found is used to evaluate~$\epsilon$ and~$\lambda_z$:
\beqa
\epsilon = x\frac{3y-1}{2a} -\frac{1}{2ax}\left(\frac{q(a^2 y-1)}{y} + 2a^2y^2 - 3y+1\right)\,, \label{epsq}\\
\lambda_z = x + a\epsilon\,. \label{lambdaq}
\eeqa
Note that~$\epsilon(y_0)$ is a well-defined number, because term~$1/x$ multiplies an expression that vanishes at that point. From~(\ref{RQq}), $\epsilon(y_0) = \sqrt{(y_0^2 + q)\Delta_y(y_0)}$.

As it was pointed out, for~$q$ that is small enough and $0<a<1$, $\lim\limits_{y\to y_{\pm}}{x^2}=+\infty$. Therefore,
\be
\lim\limits_{y\to y_{\pm}}{\epsilon}=+\infty\,,
\ee
\be
\lim\limits_{y\to y_{+}}{\lambda_z} = +\infty\,, \quad  \lim\limits_{y\to y_{-}}{\lambda_z} = -\infty\,, 
\ee
\be
\lim\limits_{y\to y_{\pm}}{\frac{\lambda_z}{\epsilon}} = \lim\limits_{y\to y_{\pm}}{\left.\frac{\lambda_z}{\epsilon}\right|_{q=0}} = -\left.\frac{1 - 3 y + a^{2} y^{2} + a^{2} y^{3} }{a y^{2} \left(1 - y \right)}\right|_{y=y_{\pm}}\,. 
\ee
In the last equation we have used the fact that the ratio does not depend on~$q$. Indeed, if we divide~(\ref{RQq}) by~$x^2$, terms that contain~$q$ vanish as~$x^2\to +\infty$ and have no impact on the value of the limit.

\printbibliography

\end{document}